\documentclass[12pt]{article}
\usepackage[latin1]{inputenc}
\usepackage[english]{babel}
\usepackage{amsfonts,amsmath,mathrsfs}
\usepackage[dvips]{graphicx}
\usepackage{natbib}
\usepackage{verbatim}
\usepackage{booktabs}
\usepackage{color}
\usepackage[english]{babel}
\usepackage{algorithmicx}
\usepackage{algpseudocode}
\usepackage{multirow}
\usepackage{algorithm}
\usepackage[top=2.5cm,left=2.8cm,right=2.8cm,bottom=2.5cm]{geometry}

\newcommand{\Real}{{\rm I}\negthinspace {\rm R}}
\newcommand{\sthat}{\widehat}
\begin{document}

\title{A note on marginal posterior simulation via higher-order tail area approximations}
\author{Erlis Ruli\footnote{Department of Statistical Sciences, Via C.\ Battisti 241, 35121 Padova, Italy; ruli@stat.unipd.it}, Nicola Sartori\footnote{Department of Statistical Sciences, Via C.\ Battisti 241, 35121 Padova, Italy; sartori@stat.unipd.it} and Laura Ventura\footnote{Department of Statistical Sciences, Via C.\ Battisti 241, 35121 Padova, Italy; ventura@stat.unipd.it}} 
%\date
\maketitle

\begin{abstract}
We explore the use of higher-order tail area approximations for Bayesian simulation. These approximations give rise to an alternative simulation scheme to MCMC for Bayesian computation of marginal posterior distributions for a scalar parameter of interest, in the presence of nuisance parameters. Its advantage over MCMC methods is that samples are drawn independently with lower computational time and the implementation requires only standard maximum likelihood routines. The method is illustrated by a genetic linkage model, a normal regression with censored data and a logistic regression model.
\end{abstract}


%%%%%%%%%%%%%%%%%%%%%%%%%%%%%%%%%%%%%%
%%%%%%%%%%%%%%%%%%%%%%%%%%%%%%%%%%%%%%

\section{Introduction}

Consider a parametric statistical model with density $f(y;\theta)$, with $\theta = (\psi,\lambda)$, $\theta\in\Theta\subseteq\Real^{d}$, where $\psi$ is a scalar parameter of interest and $\lambda$ is a $(d-1)$-dimensional nuisance parameter, and let $L(\theta)=L(\psi,\lambda)=L(\psi,\lambda;y)$ denote the likelihood function based on data $y$. Let $\pi(\theta)=\pi(\psi,\lambda)$ be a prior distribution for  $(\psi,\lambda)$ and let $\pi(\theta|y)=\pi(\psi,\lambda|y) \propto \pi(\psi,\lambda) \, \, L(\psi,\lambda)$ be the posterior distribution of $(\psi,\lambda)$. Bayesian inference on $\psi$, in the presence of the nuisance parameter $\lambda$, is based on the marginal posterior distribution
\begin{eqnarray}\label{ruli:postm}
\pi (\psi|y) \, = \, \int \pi(\psi,\lambda|y) \, d\lambda,\\\nonumber
\end{eqnarray}
which is typically computed numerically. Indeed, cumbersome numerical integration may be necessary in order to compute (\ref{ruli:postm}), in particular when $d$ is large. A variety of Markov chain Monte Carlo (MCMC) schemes have been proposed in the literature in order to approximate (\ref{ruli:postm}); see, e.g., \citet{robert2004}. All these techniques use simulation to avoid tailoring analytical work to specific models. However, these methods may present some difficulties, especially when $d$ is large, and may have poor tail behaviour. 

Parallel with these developments has been the development of analytical higher-order approximations for parametric inference in small sample (see, e.g., \citealp{brazzale2008}, and references therein).  Using higher-order asymptotics it is possible to avoid the difficulties related to numerical methods and obtain accurate approximations of (\ref{ruli:postm}) and related quantities (see, e.g., \citealp{reid96, reid03, sweeting1996}, and \citealp{brazzale07}). These methods are highly accurate in many situations, but are nevertheless underused compared to simulation-based procedures \citep{brazzale2008}. 

Starting from higher-order tail area approximations for (\ref{ruli:postm}) (see \citealp{diciccio1991, reid96, reid03} and \citealp{brazzale07}), this paper describes the implementation and the use of a sampling scheme that give rise to very accurate computation of marginal posterior densities, and related quantities, such as posterior summaries. The implementation of the proposed higher-order tail area approximation (HOTA) sampling scheme is available at little additional computation cost over simple first-order approximations, and it has the advantage over MCMC methods that samples are drawn independently in much lower computation time. This technique applies to regular models only; precise regularity conditions for their validity are given in \citet{geisser1990}.

The paper is organized as follows. Section \ref{sec:HOTA} briefly reviews higher-order approximations for the marginal posterior distribution (\ref{ruli:postm}), and for the corresponding tail area. Section 3 describes the proposed HOTA sampling scheme and its implementation. Numerical examples and applications are discussed in Section 4. Finally, some concluding remarks are given in Section 5.

%======================================================================
\section[Background]{Background on higher-order asymptotics}
\label{sec:HOTA}
%======================================================================

Let $\ell_p (\psi)=\log L(\psi,\sthat\lambda_\psi)$ be the profile loglikelihood for $\psi$, where $\sthat{\lambda}_\psi$ is the constrained maximum likelihood estimate (MLE) of $\lambda$ for fixed $\psi$. Moreover, let $\sthat{\theta}=(\sthat\psi, \sthat\lambda)$ be the full MLE, and let $j_p (\psi)=- \partial^2 \ell_p(\psi)/\partial \psi^{2}$ be the observed information corresponding to the profile loglikelihood. Standard results for partitioned matrices give $|j_p(\psi)|=|j(\psi,\sthat\lambda_\psi)|/|j_{\lambda \lambda} (\psi,\sthat\lambda_\psi)|$, where $j(\theta)=j(\psi,\lambda)$ is the observed Fisher information from $\ell(\psi,\lambda)=\log L(\psi,\lambda)$ and $j_{\lambda \lambda} (\psi,\lambda)$ is the $(\lambda,\lambda)$-block of $j(\psi,\lambda)$. The basic requirement for the approximations given in this section is that there exists an unique MLE, as occurs in many commonly used parametric models; see also \citet{geisser1990}.

The marginal posterior distribution (\ref{ruli:postm}) can be approximated with the Laplace formula (see, e.g., \citealp{tierney1986, reid96}). We get 
\begin{eqnarray}
\pi(\psi|y) \, \, \dot{=} \, \, c \,  |j_p (\sthat\psi)|^{1/2} \exp\{ \ell_p(\psi)-\ell_p(\sthat\psi) \} \, \, \frac{|j_{\lambda \lambda} (\sthat\psi, \sthat\lambda)|^{1/2}}{|j_{\lambda \lambda} (\psi, \sthat\lambda_\psi)|^{1/2}} \, \,  \frac{\pi(\psi,\sthat\lambda_\psi)}{\pi(\sthat\psi,\sthat\lambda)}
\ ,
\label{lap_1} 
\end{eqnarray}
where $c$ is the normalizing constant and the symbol ``$\dot{=}$'' indicates that the approximation is accurate to third-order, that means with error of order $O_{p}(n^{-3/2})$. Note that the approximation (\ref{lap_1}) depends on simple likelihood quantities and the prior evaluated at $(\sthat\psi,\sthat\lambda)$ and at $(\psi,\sthat\lambda_\psi)$. Then, starting from (\ref{lap_1}), an application of the tail area argument gives the corresponding third-order approximation to the marginal posterior tail area probability (see \citealp{diciccio1991, reid03}). We have
\begin{eqnarray}
\int_{-\infty}^{\psi_0} \pi (\psi|y) \, d\psi \, \,  \dot{=} \, \,   \Phi \left( r_B^{\star}(\psi_0) \right)
\ ,
\label{lap_2} 
\end{eqnarray}
where $\Phi(\cdot)$ is the standard normal distribution function, $r_p(\psi)=\mbox{sign}(\sthat\psi-\psi) [2(\ell_p(\sthat\psi) - \ell_p (\psi))]^{1/2}$ is the profile likelihood root,
\begin{equation}
q_B(\psi) = \ell_p'(\psi) \, \,  |j_p(\sthat\psi)|^{-1/2} \, \, \frac{|j_{\lambda \lambda} (\psi, \sthat\lambda_\psi)|^{1/2}}{|j_{\lambda \lambda} (\sthat\psi, \sthat\lambda)|^{1/2}} \, \,  \frac{\pi(\sthat\psi,\sthat\lambda)}{\pi(\psi,\sthat\lambda_\psi)}
\ ,
\label{ruli:qB}
\end{equation}
and
\begin{eqnarray}
r_B^{\star}(\psi)=r_p(\psi)+\frac{1}{r_p(\psi)} \log \frac{q_B(\psi)}{r_p(\psi)} 
\ .
\label{rstarb} 
\end{eqnarray}
Formula (\ref{lap_2}) gives an explicit expression for the posterior quantiles, and $s(\psi) = 1- \Phi (r^{\star}_B (\psi))$ gives the Bayesian survivor probability with third-order accuracy.  

When the particular class of matching priors is considered in (\ref{ruli:postm}) (see \citealp{tibshirani1989}),  the marginal posterior distribution for $\psi$ can be expressed as \citep{ventura2009, ventura11, venturaeal11}
\begin{eqnarray}
\pi(\psi | y) & \propto & L_{mp}(\psi) \, \, \pi_{mp}(\psi) 
\ ,
\label{post1}
\end{eqnarray}
where $L_{mp}(\psi)=L_p(\psi) M(\psi)$ is the modified profile likelihood for a suitably defined correction term $M(\psi)$ (see, among others,  \citealp[Chap. 9]{severini00} and  \citealp{pacesalvan06}), and $ \pi_{mp}(\psi) \propto i_{\psi \psi . \lambda} (\psi,\sthat\lambda_\psi)^{1/2}$ is the corresponding matching prior, with
$i_{\psi \psi . \lambda} (\psi,\lambda) = i_{\psi \psi}(\psi,\lambda) - i_{\psi \lambda}(\psi,\lambda) i_{\lambda \lambda}(\psi,\lambda)^{-1} i_{\lambda \psi} (\psi,\lambda)$ the partial information, and $i_{\psi \psi}(\psi,\lambda)$, $i_{\psi \lambda}(\psi,\lambda)$, $i_{\lambda \lambda}(\psi,\lambda)$, and $i_{\lambda \psi} (\psi,\lambda)$ blocks of the expected Fisher information $i(\psi,\lambda)$ from $\ell(\psi,\lambda)$. Accurate tail area probabilities are computable by direct integration of (\ref{post1}). In particular, in \citet{ventura11} it is shown that (\ref{lap_2}) holds with
\begin{eqnarray}
r_{B}^{\star}(\psi)=r_{p}^{\star}(\psi),
\label{tail2}
\end{eqnarray}
where $r_{p}^{\star}(\psi)$ is the modified profile likelihood root of \citet{barndorff1994}; see also \citet{cox1994inference} and \citet[Chap. 7]{severini00}. The quantity $r_{p}^{\star}(\psi)$ has the form (5) with
\begin{equation}
q_B(\psi) = \frac{\ell_p'(\psi)}{j_p(\sthat{\psi})^{1/2}} \frac{i_{\psi \psi . \lambda} (\sthat \psi,\sthat\lambda)^{1/2}}{i_{\psi \psi . \lambda} (\psi,\sthat\lambda_\psi)^{1/2}} \frac{1}{M(\psi)}
\ .
\label{tail3}
\end{equation}

%======================================================================
\section{Posterior simulation via tail area approximations}
\label{sec:algorithm}
%======================================================================
Equation (3) can be used to calculate quantiles of the marginal posterior distribution of $\psi$ but can not be used for computing moments, highest posterior density (HPD) regions, or to plot its density function. In this section the higher-order tail area approximation (HOTA) sampling scheme is presented whereby posterior summary via empirical moments, HPD, etc. is possible. Starting from (\ref{lap_2}), the simulation algorithm can be summarized as in Algorithm~\ref{alg:Marg.post}.
The quantities required for the implementation of the HOTA scheme are, apart from $\sthat\theta$ and the corresponding loglikelihood value, the profile score $\ell_{p}'(\psi)$ the constrained MLE $\sthat\lambda_{\psi}$, and the curvature of the likelihood evaluated at the constrained MLE. 

If code for the full likelihood of the model is available, constrained maximization and computation of the hessian is generally straightforward. The profile likelihood may by obtained by considering a grid of values of $\psi$ and the related loglikelihood on the constrained estimate. After that, a spline interpolator may be applied, and the profile loglikelihood and its derivative can be easily evaluated. For many statistical models, available \texttt{R} functions can be used directly for computing the full and constrained maximization, and consequently for the related profile quantities. For instance, the command \texttt{glm} of \texttt{R} can handle many generalized linear models, and it offers the option \texttt{offset} for constrained estimation. For more details on implementation issues see \citet[Chap. 9]{brazzale07}.
\begin{algorithm}[!h]
\caption[]{\label{alg:Marg.post} HOTA simulation from the marginal posterior distribution $\pi(\psi|y)$ }
\begin{algorithmic}[1]
\For {$t = 1 \to T$} 
\State generate a pseudo-random number $z \sim \mathsf{N}(0, 1)$
\State solve $r_{B}^{\star}(\psi)=z$ in $\psi$
\EndFor
\State store the vector $\psi$ as a sample from $\pi(\psi|y)$.
\end{algorithmic}
\end{algorithm}

Typically, $r_{B}^{\star}(\psi)$ is a monotonically decreasing function in $\psi$ and has a numerical discontinuity at $\sthat\psi$. In order to implement Algorithm 1, the inverse of $r_{B}^{\star}(\psi)$ is required. This may be accomplished by first applying the function $r_{B}^{\star}(\psi)$ to a grid of $\psi$ values and, secondly, smoothing its values by a spline interpolator, e.g. with the \texttt{smooth.spline} command in \texttt{R} environement. The values of $\psi$ may be set equal to the limits of the Wald confidence interval $\sthat\psi\pm 4SE(\psi)$, or its deviance version, where $SE(\psi)=j_{p}(\sthat\psi)^{-1/2}$. If the prior information is rather strong, then they can be changed in a trial-and-error fashion by looking at the $r_{B}^{\star}(\psi)$ curve. The main point is that the range of values of $r_{B}^{\star}(\psi)$ should be wide enough to include all the most probable values of the standard normal distribution, e.g. -4 and 4. Lastly, in order to avoid numerical issues in the spline smoothing step, it may be necessary to exclude values of $\psi$ close to $\sthat\psi$. For instance, during the grid construction one may exclude values of $\psi$ in $(\sthat\psi- \delta SE(\psi),\sthat\psi+\delta SE(\psi))$, for some small $\delta$; see also \citet[Chap. 9]{brazzale07}.

The HOTA simulation procedure is essentially an inverse method of sampling and it gives independent samples from (\ref{ruli:postm}) by inverting the cumulative distribution function approximation (\ref{lap_2}). In this respect, it has an obvious advantage over MCMC methods which usually requires more attention form the practitioner. Moreover, HOTA is almost automatically obtained from likelihood quantities, which opens the possibility of doing Bayesian inference with maximum likelihood routines.  We notice that a similar approach was suggested in \citet{kharroubi2010}, where the same type of approximation methods were used but through the importance sampling technique. Also their approximation methods seems to be quite accurate. Nevertheless, as far as the simulation from $\pi(\psi|y)$ is concerned, we think HOTA has an obvious advantage over the latter in that it is easier to implement and computationally faster.

%======================================================================
\section{Examples}
%======================================================================

The aim of this section is to illustrate the performance and the advantage of the HOTA sampling scheme through three examples. In all but the first example, the HOTA scheme is compared with a typical posterior approximation technique, namely the Metropolis-Hastings (M-H), which is one of the MCMC methods most widely used in practice. 

MCMC methods give autocorrelated posterior samples and, before their use, it is important to check that the chain has converged to its ergodic distribution (see, e.g., \citealp{gelman2003}). For the two examples where M-H is used, the convergence to the posterior was carefully checked by the routines of \texttt{coda} packages of \texttt{R}. As already stressed, HOTA algorithm gives independent samples. Therefore, in order to fairly compare HOTA results with the MCMC method, we need the latter to have as low autocorrelation as possible. Hence, the chains were run for a very large number of iterations, previously thinned and the initial observations discarded. After all these steps we end up with $10^{5}$ simulations in each example, with no significant autocorrelation for time lags greater than 10. 

%======================================================================
\subsection{Example 1: Genetic linkage model}
\label{ssec:ex1}
%======================================================================

We start with a simple scalar parameter problem, which has been studied also in \citet{rao1973}, \citet{dempster1977}, \citet{tanner1987} and \citet{kharroubi2010}, among others. It concerns a genetic linkage model in which $n$ individuals are distributed multinomially into four categories with cell probabilities $(\frac{1}{2}+\frac{\theta}{4}, \frac{1}{4}(1-\theta), \frac{1}{4}(1-\theta), \frac{\theta}{4})$, with $\theta\in(0,1)$. In this situation there are no nuisance parameters. Therefore, expression (\ref{lap_2}) simplifies to
\begin{eqnarray}\nonumber
\int_{-\infty}^{\theta_{0}}\,\pi(\theta|y)\,d\theta\, \dot{=}\, \Phi(r_{B}^{\star}(\theta_{0})),\\\nonumber
\end{eqnarray}
\noindent
where $r^{\star}_{B}(\theta)=r(\theta)+r(\theta)^{-1}\log\{q_{B}(\theta)/r(\theta)\}$, $q_{B}(\theta)=\ell^{\prime}(\theta)j(\sthat\theta)^{-1/2}$, $\ell^{\prime}(\theta)=d\ell(\theta)/d\theta$, and $r(\theta)=\mathrm{sign}(\sthat\theta-\theta)[2(\ell(\sthat\theta)-\ell(\theta))]^{1/2}.$

\begin{figure}[ht]\centering
\includegraphics[scale=0.5, angle=-90]{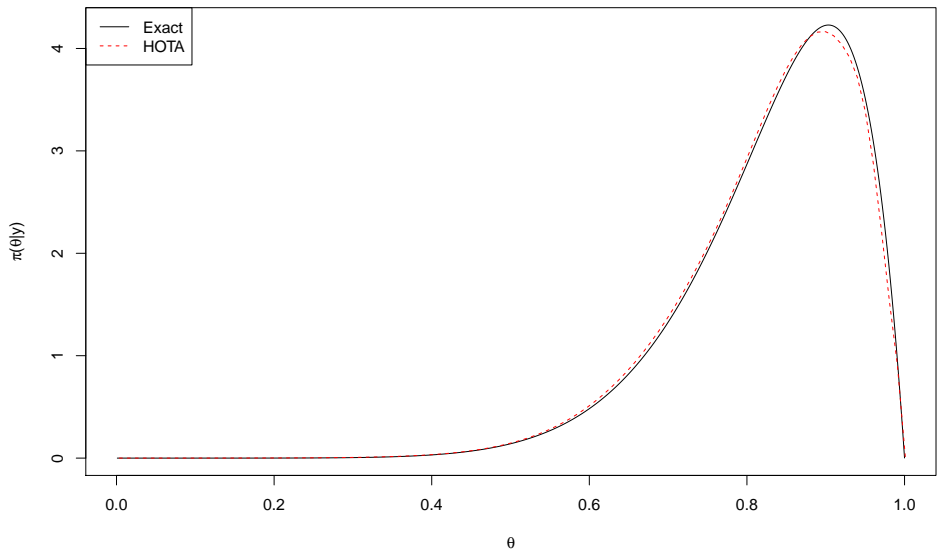}
\caption{Exact and HOTA posterior distribution in the genetic linkage model.}
\label{ruli:ex1:fig01}
\end{figure}

\citet{wei1990} and \citet{kharroubi2010} apply this model to $n=20$ animals and cell counts $y=(14,0,1,5).$ Under a uniform prior for $\theta$, the posterior density of $\theta$ is
\begin{eqnarray}\nonumber
\pi(\theta|y)\,\propto \,(2+\theta)^{14} (1-\theta)\theta^{5},\qquad \theta\in(0,1).
\end{eqnarray}
Figure~\ref{ruli:ex1:fig01} shows the posterior distribution computed with the HOTA algorithm (dashed) and the exact posterior distribution $\pi(\theta|y)$ (solid line). For the computation with HOTA algorithm, a smoothing spline was run over a grid 50 values of $\theta$ and a sample of $T=10^{5}$ points were considered. In this simple example the sample, on an Intel i5 machine with 4 GB RAM, was obtained in less than a second.

The posterior distribution appears to be extremely skewed to the right (see \citealp{tanner1987}), with a long left tail and in this case one might expect HOTA approximation to fail.  On the contrary, HOTA algorithm gets very close to the true posterior, albeit the sample size is $n=20$.

In order to further explore the behavior of the HOTA algorithm, the two posteriors are compared also in terms of some summary statistics (mean, standard deviation, 2.5 percentile, median, 97.5 percentile and 0.95 HPD credible set) in Table~1.
\begin{table}[htbp]
\centering\label{ruli:ex1:tab1}
\begin{tabular}{|l|l|l|l|l|l|l|}\hline
Posterior & Mean & St. Dev. & $Q_{0.025}$ & Median & $Q_{0.975}$& 0.95 HPD\\
\hline
Exact      & 0.831 & 0.108	&  0.57 & 0.852 & 0.978 &  (0.62, 0.994)\\
HOTA     &  0.827 &  0.109   & 0.563 & 0.848 & 0.976 & (0.617, 0.994)\\\hline
\end{tabular}
\caption{Numerical summaries of the exact and HOTA posterior distribution in the genetic linkage model.}
\end{table} 

The HOTA results are very close to those based on the true posterior.

%======================================================================
\subsection{Example 2: Censored regression}
\label{ssec:ex2}
%======================================================================

The data are taken from \citet{schmee1979} and were analyzed from a Bayesian perspective in \citet{naylor1982}, \citet{wei1990} and \citet{kharroubi2010}, among others. The data consists on temperature accelerated life tests on electrical insulation in $n=40$ motorettes. Ten motorettes were tested at each of four temperatures in degrees Centigrade ($150^{\circ}$, $17^{\circ}$, $190^{\circ}$ and $220^{\circ}$), the test termination (censoring) time being different at each temperature. 

As in \citet{schmee1979}, we consider the following model
\[
y_{i}=\beta_{0}+\beta_{1}x_{i}+\sigma\epsilon_{i},\qquad i=1,\dots,n,
\]
where $y_{i}$ is the $\log_{10}(\mathrm{failure time})$ with time in hours, $x_{i}=1000/(\mathrm{temperature}+273.2)$ and $\epsilon_{i}$ are independent standard normal errors. Reordering the data so that the first $m$ observations are uncensored, with observed log-failure times $y_{i}$, and the remaining $n-m$ are censored at times $u_{i}$, the loglikelihood function for $\theta=(\beta_{0},\beta_{1},\sigma)$ can be written as
\[
\ell(\theta)=-m\log\sigma-\frac{1}{2\sigma^{2}}\sum_{i=1}^{m}(y_{i}-\beta_{0}-\beta_{1}x_{i})^{2} + \sum_{i=m+1}^{n}\log\left\{1-\Phi\left(\frac{u_{i}-\beta_{0}-\beta_{1}x_{i}}{\sigma}\right)\right\}.
\]
In this example, for the parameter vector $(\beta_{0}, \beta_{1}, \tau)$, with $\tau=\log \sigma$, the non-informative prior $\pi(\beta_{0},\beta_{1},\tau)\propto 1$ is assumed. Clearly, the posterior distribution $\pi(\beta_{0},\beta_{1},\tau|y)$ does not have a closed form expression and direct integration is not possible in order to compute $\pi(\psi|y)$ and related quantities, where $\psi$ is one of the parameters of the model. Therefore numerical or analytical approximations are needed.

\begin{figure}[!ht]\centering
\includegraphics[scale=0.5, angle=-90]{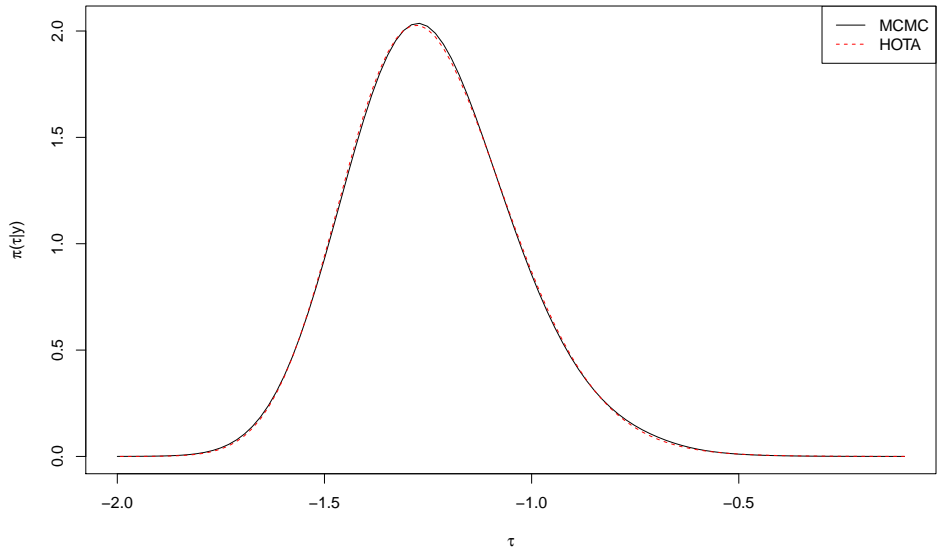}
\caption{HOTA and MCMC marginal posterior distributions of $\tau$ in the censored regression.}
\label{ruli:ex2:fig01}
\end{figure}

The marginal posterior densities $\pi(\beta_{0}|y)$, $\pi(\beta_{1}|y)$ and $\pi(\tau|y)$ were computed both with the HOTA algorithm and, for comparison purposes, with the MCMC algorithm. For the computation with HOTA, grids of 50 points were chosen for all the parameters and the total number of simulations was $T=10^{5}$. The computational time was less than 2 seconds whereas the computation with MCMC took about 95 seconds.

As an example, Figure \ref{ruli:ex2:fig01} illustrates the HOTA and MCMC marginal posterior distributions of $\tau$. Clearly, the distributions have a moderate positive asymmetry and the HOTA sampling scheme and MCMC give very similar results.

In Table~2 the two methods are compared also with respect to some summary statistics calculated over the three marginal posterior distributions. 
\begin{table}[htbp]
\centering\label{ruli:ex2:tab1}
\begin{tabular}{|l|c|l|l|l|l|l|l|}\hline
Method &Param.& Mean & St Dev. & $Q_{0.025}$ & Median & $Q_{0.975}$&0.95 HPD\\\hline

MCMC     &\multirow{2}*{$\tau$}& -1.24 & 0.201	&-1.60  &-1.253 & -0.811 & (-1.616, -0.832)\\
HOTA     &&  -1.24 &  0.202     & -1.601  & -1.251& -0.808 & (-1.624, -0.837)\\
\hline
MCMC      &\multirow{2}*{$\beta_{0}$}& -6.204 & 1.117	&	-8.57 & -6.139 &	-4.149 &( -8.413, -4.01)\\
HOTA     &&  -6.191 &  1.128     & -8.596  & -6.134 & -4.13 &(-8.475, -4.038)\\
\hline
MCMC      &\multirow{2}*{$\beta_{1}$}& 4.409 & 0.518	&3.461 & 4.382 &	5.512& (3.425, 5.47)\\
HOTA     &&  4.401 &  0.521     & 3.459 & 4.37 & 5.521& (3.398, 5.443)\\
\hline
\end{tabular}
\caption{Numerical summaries of the MCMC and HOTA marginal posterior distributions in the censored regression.}
\end{table} 

The results based of the two methods are in good agreement, and this holds also for the 0.95 HPD credible sets.

%======================================================================
\subsection{Example 3: Logistic regression}
\label{ssec:ex3}
%======================================================================

In this example we consider a logistic regression model applied to the \texttt{urine} dataset reported in  \citet{andrews1985}; see also \citet[Chap. 4]{brazzale07}. This dataset concerns calcium oxalate crystals in samples of urine. The binary response $y$ is an indicator of the presence of such crystals, and there are six explanatory variables: specific gravity (\texttt{gravity}), i.e. the density of urine relative to water; pH (\texttt{ph}); osmolarity (\texttt{osmo}, mOsm); conductivity (\texttt{conduct}, mMho); urea concentration (\texttt{urea}, millimoles per litre); and calcium concentration (\texttt{calc}, millimoles per litre). After dropping the two incomplete cases, the dataset consists of 77 observations.

Let us denote by $X$ the $(n\times7)$ design matrix composed by a vector of ones and the six covariates, and let $\beta=(\beta_{0},\dots,\beta_{6})$ be the vector of parameters, with $\beta_{0}$ being the intercept. The loglikelihood function for $\beta$ is
\[
\ell(\beta)= y^{T}X\beta-\sum_{i=1}^{n}\log\{1+\exp\{x_{i}^{T}\beta\}\},\]
where $x_{i}$ represents the the $i$-th row of $X$ and $y$ is the vector of binary responses. For illustrative purposes let us consider the non-informative prior $\pi(\beta)\propto 1$, which leads to a posterior distribution being proportional to the likelihood.

\begin{figure}[!htb]\centering
\includegraphics[scale=0.5, angle=-90]{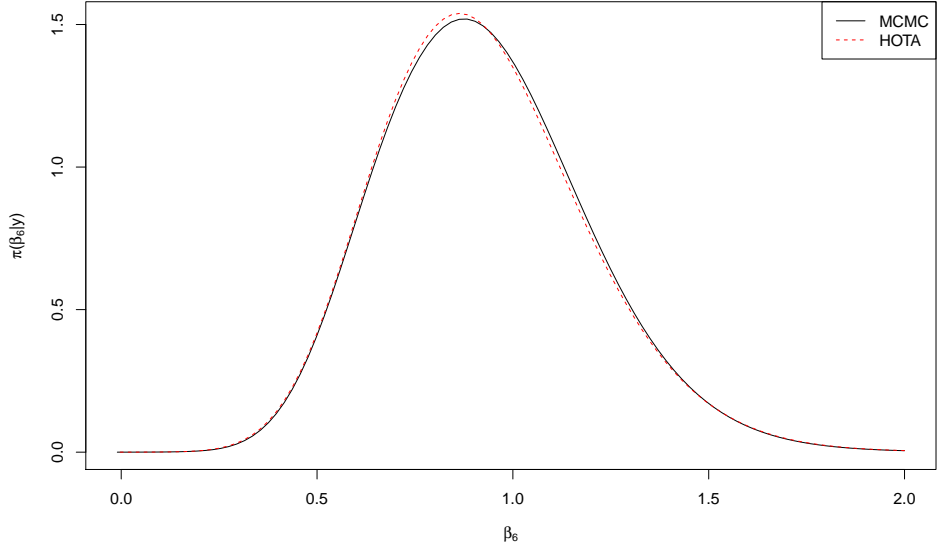}
\caption{HOTA and MCMC marginal posterior distributions of $\beta_{6}$ in the logistic regression.}
\label{ruli:ex3:fig01}
\end{figure}

From a computational point of view, as far as HOTA is concerned,  a grid of 50 points were used and a sample of $T=10^{5}$ values were simulated from each marginal posterior in about 5 seconds. Whereas, the computation time with MCMC was about 100 seconds. The comparison of the HOTA algorithm with MCMC is illustrated for the marginal posterior of $\beta_{6}$ in Figure~\ref{ruli:ex3:fig01}. 
The posterior distribution of $\beta_{6}$ appears to be slightly skewed. Again HOTA and MCMC produce very similar results. 

Let us further consider the posterior distributions of the last three parameters, namely $\beta_{4}$, $\beta_{5}$ and $\beta_{6}$,  over which we calculate the summary statistics considered also in the previous examples. These parameters represent the effects of conductivity, urea concentration and calcium concentration respectively. From a frequentist point of view, \citet{brazzale07} found that the two tailed p-values associated to these parameters were respectively 0.08493, 0.04703 and 0.00121 and, under frequentist higher order refinements, only $\beta_{6}$ was significantly different from zero at 5\% significance level. Bayesian analysis, instead, suggests that with 0.95 probability the first two variables have a negative effect whereas the last one has a positive effect (see the 0.95 HPDs in Table~3). Moreover, with 0.99 probability only the HPD for calcium concentration does not include zero.
\begin{table}[htbp]
\centering\label{ruli:ex3:tab1}
\begin{tabular}{|l|c|l|l|l|l|l|l|}
\hline
Method &Param.& Mean & St Dev. & $Q_{0.025}$ & Median & $Q_{0.975}$&0.95 HPD\\
\hline
MCMC      &\multirow{2}*{$\beta_{4}$}& -0.536 & 0.277	&-1.108  &-0.525 & -0.025 &(-1.084, -0.005)\\
HOTA     &&  -0.546 &  0.281     & -1.128  & -0.535 & -0.026 & (-1.108, -0.01)\\
\hline
MCMC      &\multirow{2}*{$\beta_{5}$}& -0.039 & 0.018	&	-0.076 & -0.038 &	-0.006 & (-0.075, -0.005)\\
HOTA     &&  -0.039 &  0.018     & -0.077  & -0.039 & -0.006 &(-0.076, -0.005)\\
\hline
MCMC      &\multirow{2}*{$\beta_{6}$}& 0.935 & 0.268	&0.471 & 0.914 &	1.524&(0.43, 1.464)\\
HOTA     &&  0.926 &  0.267     & 0.466 & 0.904 & 1.509&(0.429, 1.459)\\
\hline
\end{tabular}
\caption{Numerical summaries of the MCMC and HOTA marginal posterior distributions for $\beta_{4}$, $\beta_{5}$ and $\beta_{6}$ in the logistic regression.}
\end{table}

Another advantage of the HOTA simulation scheme is that sensitivity analyses with respect to the prior can be done very quickly. In fact, for a different prior distribution, say $\pi_{1}(\theta)$, the values of $\sthat\theta$ and $\sthat\theta_{\psi}$ remain the same and only the prior ratio in equations (\ref{ruli:qB}) and (\ref{rstarb}) has to be recomputed. Moreover, using the same set of normal random numbers one can compare the prior's effect on the posterior distribution under the same Monte Carlo variation.

As an illustrative example we compare the marginal posterior distribution for $\beta_{6}$ under the matching prior $\pi_{mp}(\psi)$ and the multivariate normal prior with independent components. For the logistic regression model and in the case of the matching prior (see Sec. \ref{sec:HOTA}), equation (\ref{tail3}) becomes
\begin{eqnarray}
\nonumber
q_{B}(\psi)&=&(\sthat\psi-\psi)\frac{|j_{\lambda \lambda} (\psi, \sthat\lambda_\psi)|^{1/2}}{|j_{\lambda \lambda} (\sthat\psi, \sthat\lambda)|^{1/2}} \,j_{p}(\sthat\psi)^{1/2}.
\end{eqnarray}
Whereas the multivariate normal prior, very commonly used in practice, can be written as $\pi(\beta) \sim \mathrm{N}_{d}(\mu_{0}, \Sigma_{0})$. For illustration purposes we take $\mu_{0}=0_{d}$ where $0_{d}$ is a $(1\times d)$ vector of zeros and $\Sigma_{0}=kI_{d}$, with $I_{d}$ the $(d\times d)$ identity matrix. In our case $d=7$ and we consider three possible alternatives for $k$, namely 10, 35 and 100.
\begin{figure}[!htb]\centering
\includegraphics[scale=0.5, angle=-90]{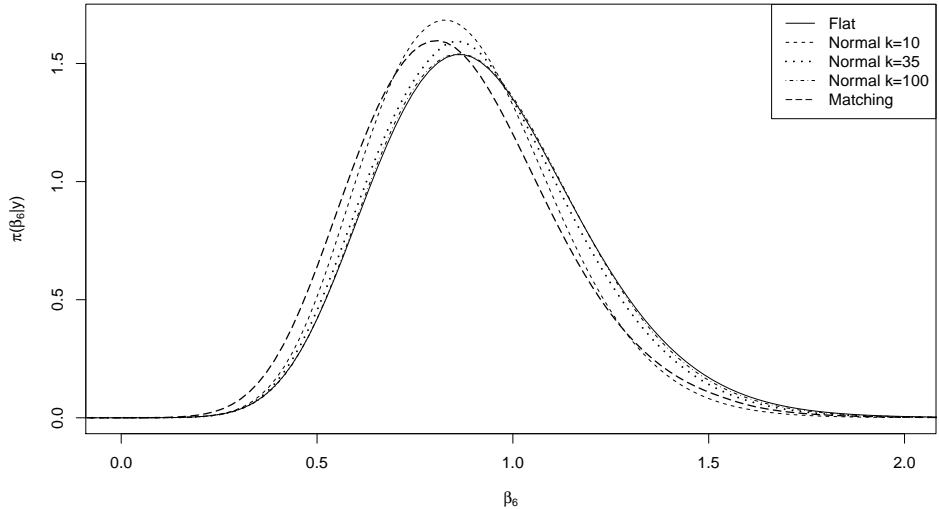}\\
\caption{HOTA marginal posteriors of $\beta_{6}$ with the flat (solid line), normal $k=10$ (dashed), normal $k=35$ (dotted), normal $k=100$ (dash-dotted) and the matching (long dashed) prior.}
\label{ruli:ex3:fig02}
\end{figure}
Figure \ref{ruli:ex3:fig02} illustrates the marginal posterior distributions for $\beta_{6}$ with the four different priors along with the flat prior, all computed both with HOTA. Lastly, recall that the flat prior can also be seen as a normal prior with zero mean and infinite variance, e.g. $k=\infty$.
\begin{figure}[!htb]\centering
\includegraphics[scale=0.5, angle=-90]{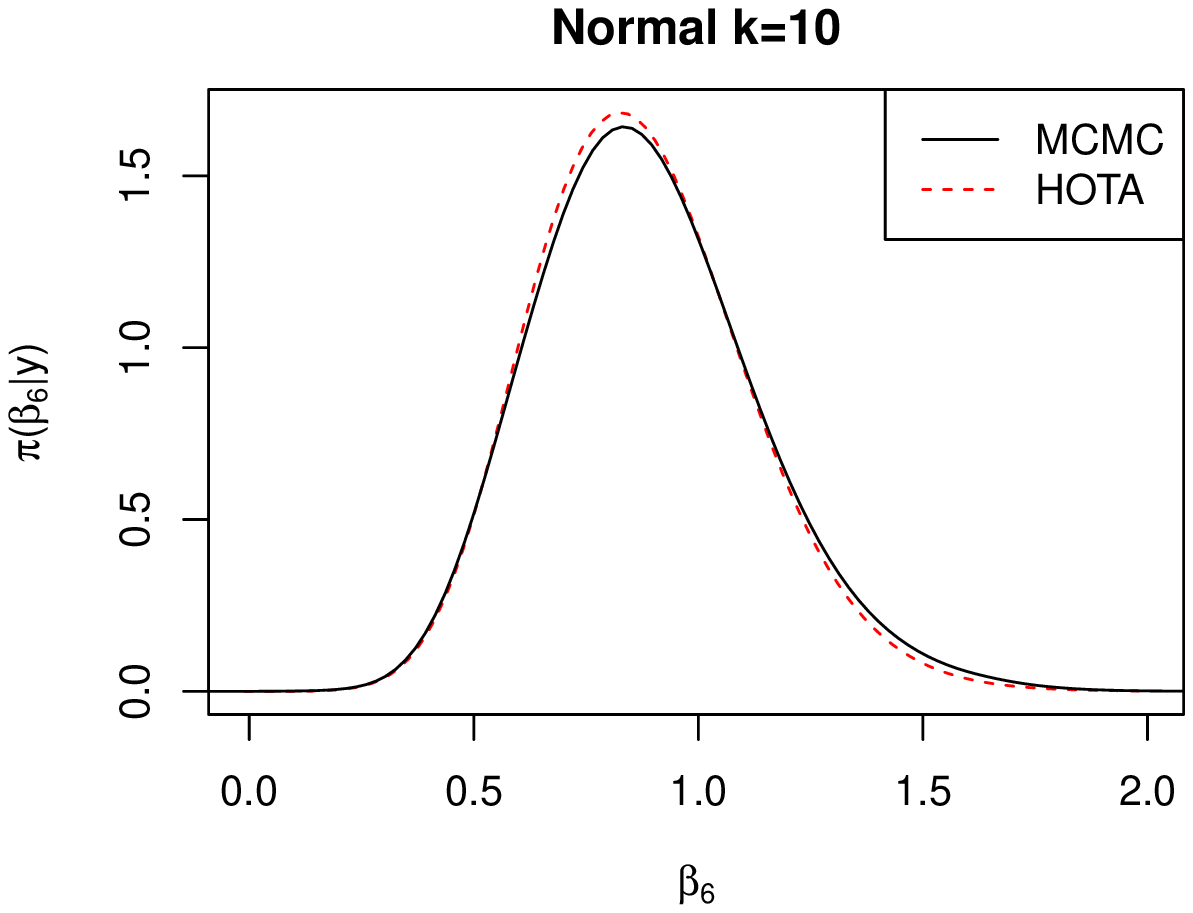}
\includegraphics[scale=0.5, angle=-90]{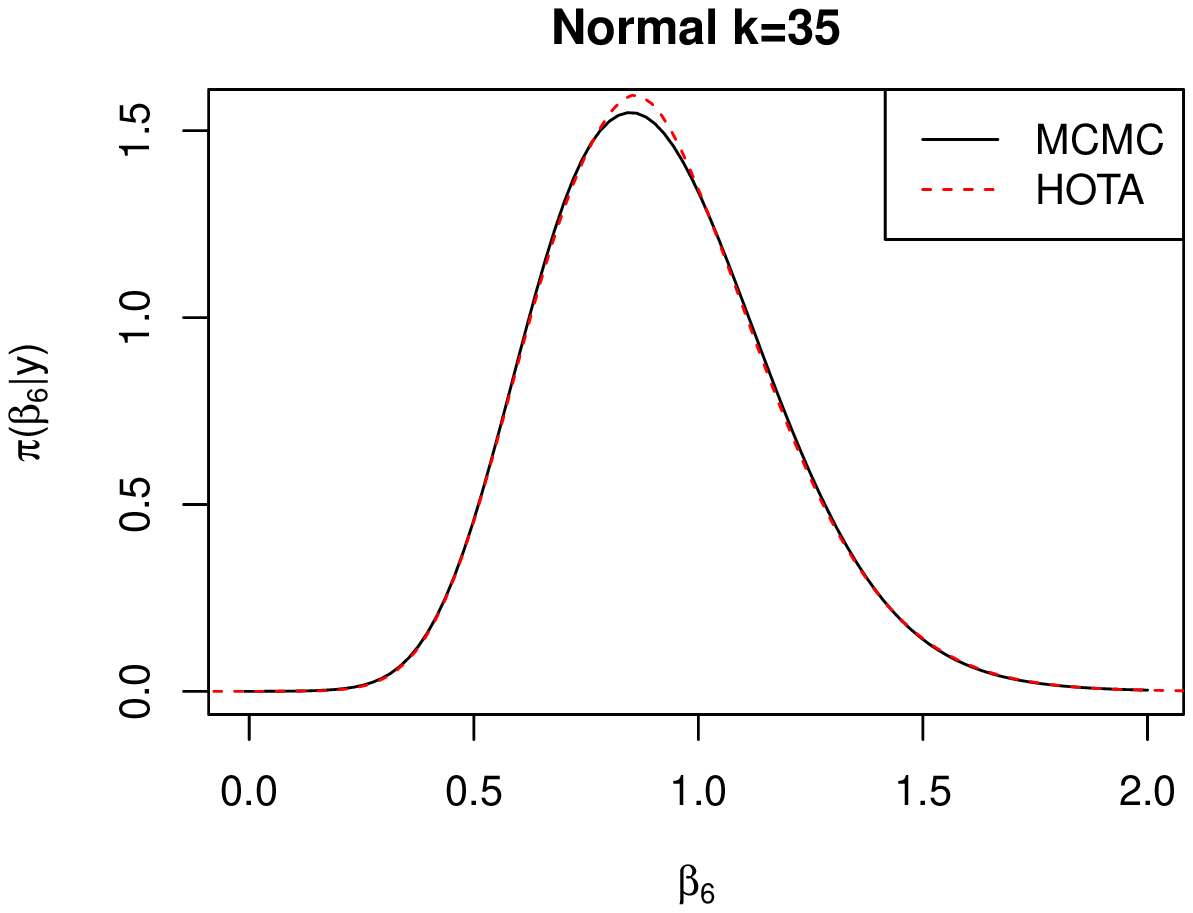}\\
\includegraphics[scale=0.5, angle=-90]{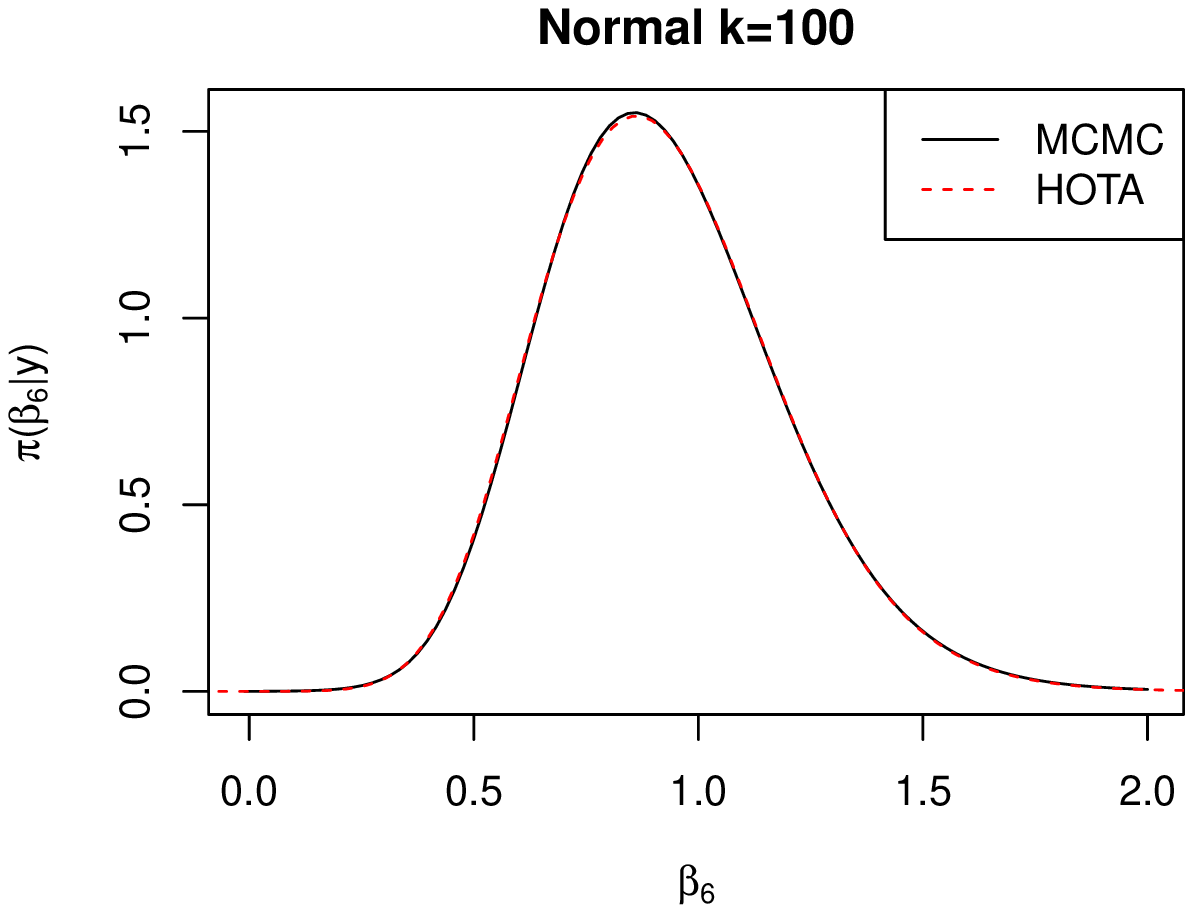}
\includegraphics[scale=0.5, angle=-90]{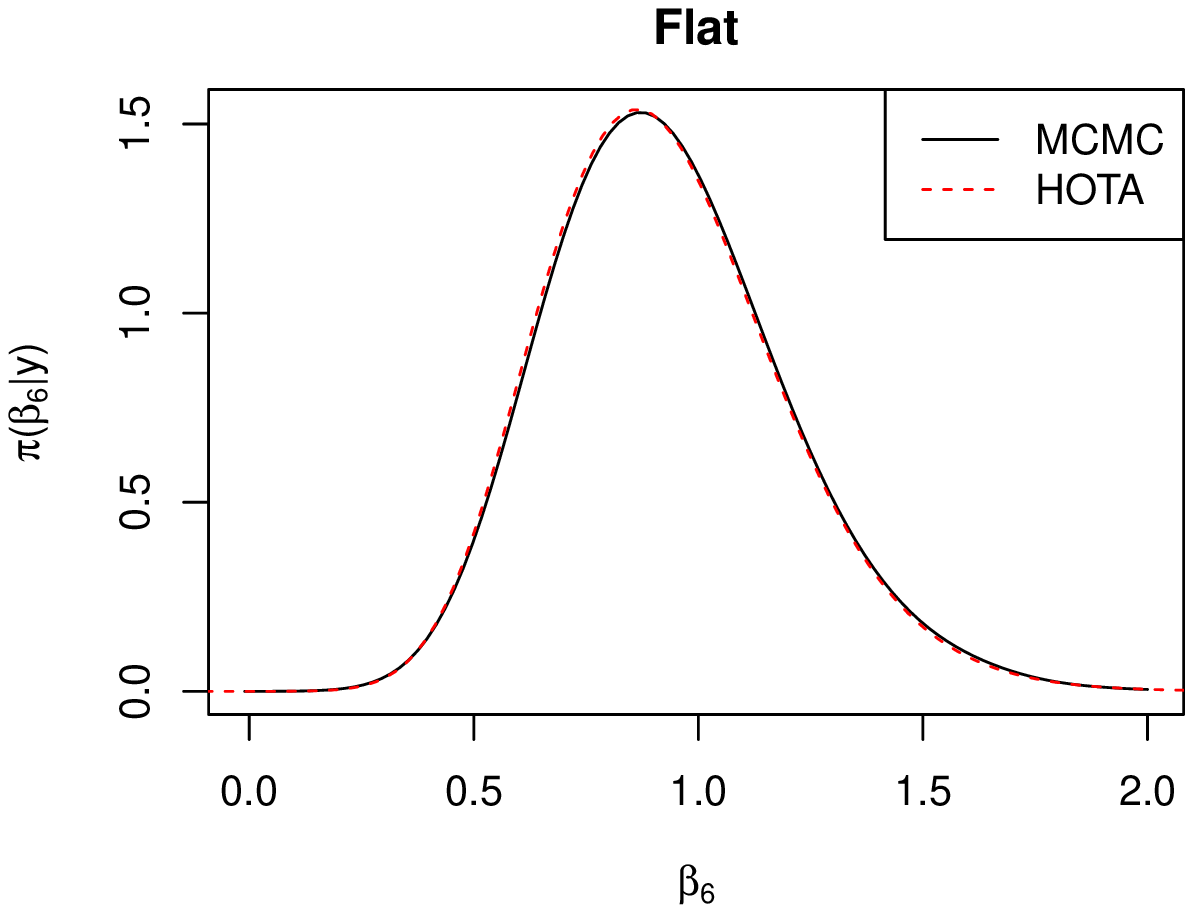}
\caption{HOTA and MCMC marginal posteriors of $\beta_{6}$ with normal $k=10$, normal $k=35$, normal $k=100$ and the flat prior.}
\label{ruli:ex3:fig03}
\end{figure}
As expected, the posterior distribution with lower $k$ tends to be more concentrated and closer to zero, whereas as $k$ increases the influence of the prior diminishes and the posterior is dominated by the data. In fact for $k=100$ and $k=\infty$, e.g. the flat prior, the two posteriors are almost indistinguishable. Interestingly, also the matching prior appears to be quite far from the non-informative flat prior and closer to the normal prior with $k=10$. 

The comparison between MCMC and HOTA marginal posterior distributions for $\beta_{6}$ is illustrated in Figure~\ref{ruli:ex3:fig03}. Since the matching prior is defined for a scalar parameter only, e.g. $\pi_{mp}(\beta_{6})$, this advantage in terms of elicitation becomes a problem for MCMC. Therefore in this case we avoid the comparison of MCMC with HOTA. We notice a slight discrepancy in the agreement between HOTA and MCMC approximation as $k$ gets lower. We will discuss more on this point in Section\ref{sec:conclusion}.

As a numeric check, Table~4 gives some summary statistics of the marginal posterior distributions computed with HOTA and MCMC for the four alternative prior distributions, namely the matching prior and the multivariate normal prior with different values of $k$.
\begin{table}[htbp]
\footnotesize
\centering\label{ruli:ex3:tab2}
\begin{tabular}{|l|l|c|l|l|l|l|l|l|}
\hline
Prior &Method&Param.& Mean & St Dev. & $Q_{0.025}$ & Median & $Q_{0.975}$&0.95 HPD\\
\hline
Matching     &HOTA  &\multirow{7}*{$\beta_{4}$}&  -0.508 &0.273 &-1.08    &-0.496  & -0.007 & (-1.048,  0.02)\\
N, $k=10$  &MCMC &&  -0.59   &0.257 &-1.123  &-0.579  &-0.117  &(-1.092, -0.093) \\
N, $k=10$  &HOTA  &&  -0.601 &0.243 & -1.095 & -0.593 & -0.145 & (-1.094, -0.144) \\
N, $k=35$  &MCMC &&  -0.555 &0.266 & -1.107 & -0.545 & -0.061 & (-1.088, -0.046) \\
N, $k=35$  &HOTA  && -0.564  &0.27   &  -1.123& -0.554 & -0.062 & (-1.106, -0.049)\\
N, $k=100$&MCMC &&-0.544   &0.273 & -1.107& -0.532 & -0.039 & (-1.088, -0.026)\\
N, $k=100$&HOTA  &&  -0.554 &0.276 & -1.127& -0.543 &-0.046  & (-1.107, -0.029)\\
\hline
Matching              &HOTA  &\multirow{7}*{$\beta_{5}$}& -0.037 &  0.018 & -0.074 & -0.037& -0.005 & (-0.072, -0.003) \\
N, $k=10$   &MCMC && -0.04   & 0.017  & -0.075 & -0.04 &-0.009 & (-0.074, -0.008)\\
N, $k=10$   &HOTA  && -0.041 & 0.016  & -0.074 & -0.04 & -0.011 & (-0.073 -0.01)\\
N, $k=35$   &MCMC && -0.04   & 0.017  & -0.075 & -0.039 & -0.008 & (-0.074, -0.007)\\
N, $k=35$   &HOTA  && -0.04   & 0.017  & -0.076 & -0.039 & -0.007 & (-0.075, -0.007)\\
N, $k=100$ &MCMC && -0.039 & 0.018 &-0.075 &-0.039  &-0.007 &(-0.075, -0.006)\\
N, $k=100$ &HOTA  && -0.04   & 0.018  & -0.077 &-0.039 &-0.007 &(-0.075, -0.006)\\
\hline
Matching              &HOTA  &\multirow{7}*{$\beta_{6}$}&  0.862  &0.257  &0.419  &0.841  &1.422  &(0.375,  1.365)\\
N, $k=10$   &MCMC &&  0.884  &0.25   &0.451  &0.866  &1.426  &(0.433,  1.401)\\
N, $k=10$   &HOTA  &&  0.873  &0.239  &0.454  &0.858  &1.384  &(0.43,  1.353)\\
N, $k=35$   &MCMC &&  0.909  &0.259  &0.458  &0.889  &1.468  &(0.427,  1.425)\\
N, $k=35$   &HOTA  & &  0.909  &0.258  &0.462  &0.889  &1.468  &(0.44,  1.436)\\
N, $k=100$ &MCMC && 0.924  &0.264  &0.468  &0.902  & 1.496  &(0.435,  1.45)\\
N, $k=100$ &HOTA  && 0.922  &0.263  &0.468  &0.901  &1.493  &(0.435,  1.445)\\
\hline
\end{tabular}
\caption{Numerical summaries of the marginal posteriors of $\beta_{4}$, $\beta_{5}$ and $\beta_{6}$ with the normal prior (N) and the matching prior based on MCMC and the HOTA approximation.}
\end{table} 

We note that the marginal posterior for $\beta_{5}$ seems quite robust with respect to the prior specification and HOTA and MCMC are in a stunning agreement. Whereas for $\beta_{4}$ and $\beta_{6}$ there are some minor differences. For instance, with the matching prior the 0.95 HPD of $\beta_{4}$ includes the zero value, whereas the 0.95 HPDs based on the other two marginal posteriors obtained with the same prior do not include the zero value. The agreement between HOTA and MCMC remains still at an acceptable level. 
%======================================================================
\section{Concluding remarks}
\label{sec:conclusion}
%======================================================================
By considering higher-order tail area approximations of a marginal posterior distribution for a scalar parameter of interest, we illustrate the HOTA algorithm for simulating independent samples from the posterior distribution. The HOTA algorithm is essentially based on standard likelihood quantities, and it opens the possibility of doing Bayesian inference with usual maximum likelihood routines' output and prior distributions. In cases where it can be applied, the HOTA scheme has the advantage over MCMC methods that it does not need convergence check and simulations are obtained almost in an automatic way, without having to rely on proposal distributions (like in M-H), importance function (importance sampling), full conditional posterior distribution (Gibbs sampling) or latent structure representation of the posterior (data augmentation). 

The  HOTA sampling scheme may be a very useful tool in the context of Bayesian robustness with respect to the prior distribution (see \citealp{kass1989}). Moreover, with the HOTA algorithm prior's effect on the posterior distribution can be appreciated with the Monte Carlo variation being the same by just using the same set of normal random numbers.  Sometimes, repeated sampling properties of the posterior distribution are needed, and also in this respect HOTA may be very useful because of its reduced computation time. Finally, in the examples considered in this article, the algorithm appears to be accurate also for small sample sizes.

As underlined in Section~\ref{ssec:ex3}, informative priors may imply a slight deterioration in HOTA's accuracy. This is not surprising. Indeed, HOTA is essentially based on the MLE and if the prior information is large, then the MLE and the posterior mode could be substantially different and obviously approximating the posterior distribution around the MLE could be misleading. An instance of this situation happens if one uses the Zellner's G-prior (see \citealp[p. 107]{marin2007}) in the logistic regression of Section~\ref{ssec:ex3}. In fact, in this case the prior distribution puts a considerable mass on zero and tends to be flat elsewhere. Therefore, the posterior mode and MLE are remarkably different and, the HOTA's accuracy slightly deteriorates. Clearly, this is due also to the moderate sample size. 

The problem of accuracy when posterior mode and MLE are different is already known in the literature (see \citealt[Chap. 11]{davison2003}) and we stress that it is a minor issue for at least three reasons. First, one can incorporate the prior in the likelihood function and consider the Laplace  expansion on the posterior mode instead of the MLE. \citet[p. 600]{davison2003} gives the expression of the tail area probability in this case and he claims that this approximation tends to be more accurate than the usual one. Second, priors such as the G-prior were essentially suggested as non-informative priors for model choice, which is a topic we are not addressing here. Nevertheless, in Bayesian estimation the use of flat priors or proper priors, perhaps with large variance, is a more usual practice; see for instance \citet{gelman2003}. Lastly, since the HOTA simulation scheme is based on a third-order tail area approximation we expect its accuracy to increase with sample size.

The proposed procedure can be applied only for marginal posterior distributions for scalar components. How to extend it to multidimensional marginal posterior distributions is an open question. This is not a trivial problem since HOTA is based on the higher-order approximation (3) which, at present, does not exist for dimension larger than one.  

%======================================================================
%\appendix
%\section*{Appendix}
%======================================================================

%======================================================================
\begin{description}
\item[Acknowledgement]This work was supported by a grant from the University of Padua (Progetti di Ricerca di Ateneo 2011) and by the Cariparo Foundation Excellence-grant 2011/2012.
\end{description}
%======================================================================

\bibliographystyle{chicago}
\bibliography{main}
%======================================================================

\end{document}